\documentclass{article}
\usepackage{smc2021}
\usepackage[caption=false, font=footnotesize]{subfig}
\usepackage{paralist}
\usepackage[figure,table]{hypcap}
\usepackage{multirow}

\usepackage[whole]{bxcjkjatype}

\usepackage{alphabeta}
\usepackage{arabtex}
\usepackage[LFE,LAE,LGR,T2A,T1]{fontenc}
\usepackage[greek, russian, main=english]{babel}

\def\papertitle{Exploring modality-agnostic representations for music classification}

\author[1]{\mbox{\firstname{Ho-Hsiang}\lastname{Wu}}}
\author[1,2]{\mbox{\firstname{Magdalena}\lastname{Fuentes}}}
\author[1,2]{\mbox{\firstname{Juan}\middlename{Pablo}\lastname{Bello}}}

\affil[1]{\department{Music and Audio Research Laboratory}\institution{New York University}\city{New York}\state{NY}\country{USA}\affiliationtype{University}}
\affil[2]{\department{Center for Urban Science and Progress}\institution{New York University}\city{New York}\state{NY}\country{USA}\affiliationtype{University}}

\completesetup

\title{\papertitle}
\begin{document}
\capstartfalse
\maketitle
\capstarttrue

\begin{abstract}
    Music information is often conveyed or recorded across multiple data modalities including but not limited to audio, images, text and scores. However, music information retrieval research has almost exclusively focused on single modality recognition, requiring development of separate models for each modality. Some multi-modal works require multiple coexisting modalities given to the model as inputs, constraining the use of these models to the few cases where data from all modalities are available. To the best of our knowledge, no existing model has the ability to take inputs from varying modalities, e.g. images or sounds, and classify them into unified music categories. We explore the use of cross-modal retrieval as a pretext task to learn modality-agnostic representations, which can then be used as inputs to classifiers that are independent of modality. We select instrument classification as an example task for our study as both visual and audio components provide relevant semantic information. We train music instrument classifiers that can take both images or sounds as input, and perform comparably to sound-only or image-only classifiers. Furthermore, we explore the case when there is limited labeled data for a given modality, and the impact in performance by using labeled data from other modalities. We are able to achieve almost 70\% of best performing system in a zero-shot setting. We provide a detailed analysis of experimental results to understand the potential and limitations of the approach, and discuss future steps towards modality-agnostic classifiers.
\end{abstract}

\section{Introduction}\label{sec:introduction}

Musical objects and concepts appear in different heterogeneous data modalities, including but not limited to audio, images, text and scores, where sonic, visual and tactile modalities contribute to the overall experience. However, most music information retrieval (MIR) research has largely focused on developing systems that interact with a single modality, requiring development of separate models for audio, image or text, and over simplifying the musical modeling. There are approaches that exploit multiple modalities \cite{baltruvsaitis2018multimodal, xu2015show, fang2015captions, cramer2019look}, but existing multi-modal systems in the context of MIR require \textit{coexisting} modalities as inputs \cite{yazdani2011affective, schindler2015audio, slizovskaia2017musical, oramas2018multimodal, duan2018audiovisual, choi2019zero}, which is a big constrain for their deployment since it limits the scope of systems to only work when the modality they have been design for is at hand. 

In a context of rapidly increasing availability of information in all forms (video, audio, text, etc) it is desirable that models are able to overcome this single-modality limitation and can interact with information in any common form, for instance, a system able to classify musical instruments by the way they look and sound. To the best of our knowledge, no existing model in the context of MIR can be used if one of those modalities is missing (e.g. if it was trained with audio and text, can not be used in a dataset with only audio).


Based on recent work \cite{arandjelovic2018objects} we hypothesize that \textit{modality-agnostic} systems can be developed by learning joint representations from different modalities when they represent the same concepts. If the embedding of an image of a guitar and the sound of a guitar are similar to each other (i.e. grouped closely in the embedding space) but different from those of a piano, we can build classification systems that would work with either image or audio. This would allow to train models in settings where there is big amounts of data from one modality but not from the other, but still be able to work in both cases. 

This type of approach, often called \textit{translation} since it implies "translating" one modality to another (e.g. being able to retrieve an image with a description of it) has received renewed attention recently given the combined efforts of the computer vision and natural language processing communities, and has been gaining more interests in the MIR community \cite{dorfer2018learning, zhang2018siamese, lee2019learning, li2019query, watanabe2019query, zalkow2020learning, arandjelovic2017look}. Recently, it has been proposed to learn translated representations using self-supervision  \cite{arandjelovic2018objects} which is very promising since it doesn't rely on human-annotated data, but has the drawback of requiring millions pairs of raw data to train embedding models from scratch. We propose an intermediate solution, to use pre-trained embeddings and only learn the translation between them in a self-supervised manner, as a way of relaxing the amount of computation time and data needed for training the system. 

In this paper, we take the first steps towards modality-agnostic music classification. We focus on the problem of classifying musical instruments using audio and/or image. We  investigate the use of pre-trained audio and image embeddings in combination with training translation models to obtain a joint representation, in a self-supervised setting. We use the learned representations to train modality-agnostic classifiers in a supervised manner, and we investigate the performance of the classifier compared to its single-modality counterpart in different scenarios, including one with varying amount of data available from either modality. Our implementation is available in \url{https://github.com/hohsiangwu/crossmodal}.



\section{Method}\label{sec:method}

\begin{figure}[ht]
\centering
\includegraphics[width=\linewidth]{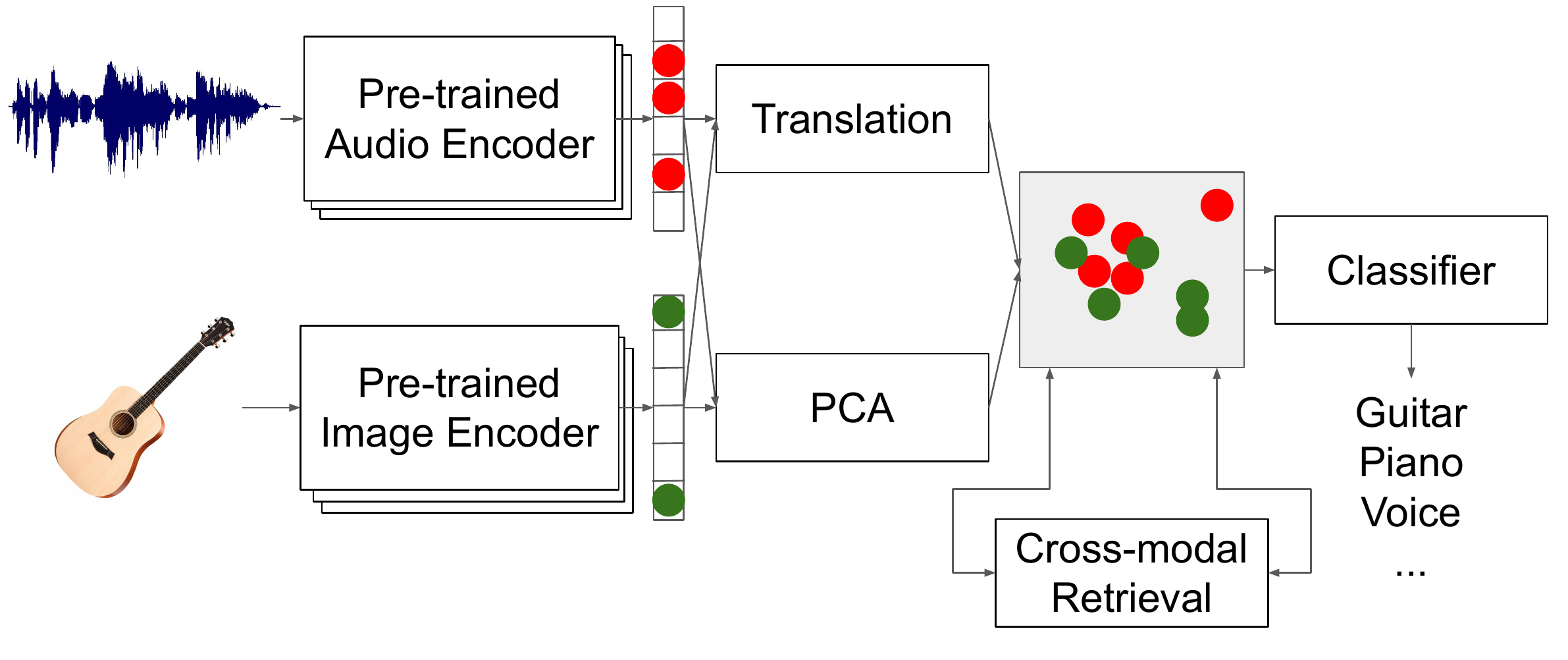}
\caption{Method overview. The pre-trained audio and image embeddings are projected to a joint space by either the translation model or PCA, and the obtained embeddings are used to train the classifier in the downstream task: musical instrument classification. Cross-modal retrieval is used to select best configuration of pre-trained embeddings.}
\label{fig:system}
\end{figure}

Our method is summarized in Figure \ref{fig:system}. It consists of three different stages: 1) First, we select a set of pre-trained embeddings from both audio and image, and translate or project the pre-trained embeddings into a common space, either by training a translation model, or simply using principal component analysis (PCA) to convert both embeddings to the same dimension; 2) We then conduct a study to find the best combination by comparing configuration performances in cross-modal retrieval; and 3) We use the resulting joint embeddings to train a classifier in a supervised setting and study the performance of the different configurations (i.e. translation vs. PCA vs. single modality) with different amount of data from each modality. We explain the different stages of the method in the following.

\subsection{Pre-trained Image and Audio Embeddings}
We select a set of state-of-the-art embeddings for both image and audio. For image embeddings, we use two pre-trained embedding models provided by \textit{keras} library\footnote{\url{https://keras.io/api/applications/}},  trained using the ImageNet \cite{deng2009imagenet} dataset on a  classification task. In particular, we use 
VGG Net \cite{simonyan2014very} and ResNet \cite{he2016deep}. These are both deep convolutional neural networks based architectures. We refer to \cite{simonyan2014very, he2016deep} for further details. For both models, we remove the last layer and apply average pooling to get the final image embeddings.

We also use pre-trained models to obtain the audio embeddings, particularly VGGish \cite{hershey2017cnn}
, and YamNet. Finally, we use the open source implementation of OpenL3\footnote{\url{https://github.com/marl/openl3}} \cite{cramer2019look} trained with music data from AudioSet \cite{audioset} to obtain another pair of  image and audio embeddings.


\begin{table}[h]
\centering
\begin{tabular}{c|c|c}
\hline
 Embedding model & \# Parameters & Output dimension \\
\hline
OpenL3 (Image) & 4.7M & 8192 \\
VGG16 & 15M & 512 \\
ResNet50 & 23.6M & 2048 \\
OpenL3 (Audio) & 9M & 6144 \\
VGGish & 62M & 128 \\ 
YamNet & 3.2M & 1024 \\
\end{tabular}
\caption{Overview of pre-trained image and audio embedding models.}
\label{tab:image_audio_model}
\end{table}




In Table \ref{tab:image_audio_model} we summarize the characteristics of each image and audio embedding model. Pre-trained VGG and ResNet image embeddings, VGGish and YamNet audio embeddings are trained on classification tasks, while OpenL3 is trained with audio-visual correspondence without labeled data.

To select the best combination, we evaluate how good the different pairs of embeddings blend together in a common space using a translation model. To quantify the success of this translation, we perform cross-modal retrieval (i.e. retrieve the image of an instrument using it's respective sound and vice versa) as further explained in Section \ref{ssec:cross-modal-retrieval}. Our reasoning behind this is that for the modality-agnostic classifier to be successful, the embeddings should be very close to each other in the joint embedding space, and so they should be accurate in a cross-modal retrieval task when retrieving examples by distance. We select the best performing pair of audio and image embeddings and use it for the following stages.

\subsection{Translation and Dimensional Reduction}


We explore two ways of relating the audio and image embeddings: translation and a simple dimensional matching with PCA. For translation, in the self-supervised learning literature, various metric learning losses are used to learn a shared embedding space \cite{chopra2005learning, hadsell2006dimensionality, wang2017deep}. In particular, the \textit{Contrastive loss} \cite{gutmann2010noise, oord2018representation} works well empirically with a careful selection of negative samples. It aims to minimize the distance of a given sample to positive examples (i.e. samples semantically related) while increasing the distance to negative examples concurrently. For the translation layer, we implement a 2 layer multi-layer perceptron (MLP) network, with pre-trained image and audio embeddings as inputs and train using contrastive loss, with cosine distance. 
We train the translation model using sample pairs from both modalities \textit{without labels}, in particular we use the Musical Instruments AudioSet subset, as explained in Section \ref{sec:experi}. The output dimension is 128 for all of our embeddings.

As baseline, we apply PCA to each pre-trained embedding model to reduce their dimension to 128, and we train a single-modality classifier as well as a multi-modality classifier with such embeddings. The idea is to understand whether a simple solution is enough to build a modality-agnostic classification system, where the classifier is mainly responsible for the work of translating the modalities and learning the mapping to the labels.

\subsection{Classification}\label{sec:multimodal_classification}

We work with random forest classifiers. We train multi-modal (MM) classifiers either with translation (MMT) or PCA (MMP) and single-modality (SM) classifiers using dimension-reduced embeddings from audio (SMA) or image (SMI). We study how translation affects performance in scenarios with different amount of data used to train the classifiers. We do so by training the MMT and MMP with data from one modality and testing in the other, which we call \textit{target modality}. We incorporate data from the target modality to the training of the classifiers by batches and see the impact in performance.

\begin{figure*}[ht!]
\centering
\includegraphics[width=\linewidth]{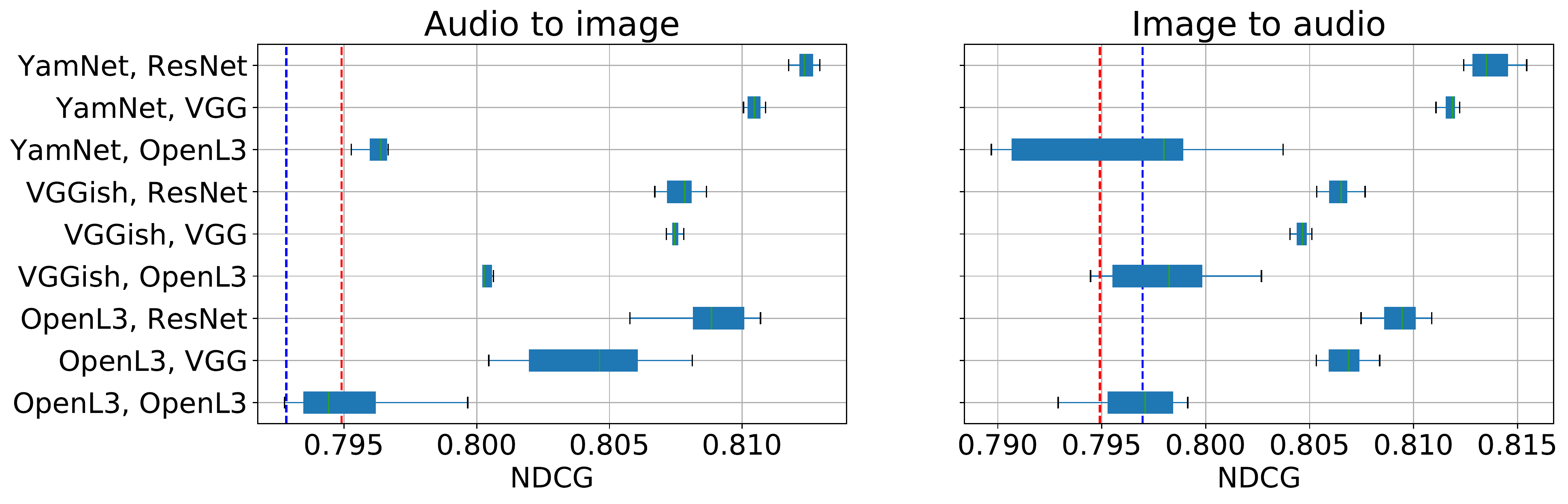}
\caption{Cross-modal retrieval results with NDCG scores on the x-axis. On the y-axis we have different combinations of \textit{audio, image} pre-trained embeddings used to train the translation model. The red dotted line is a random baseline, and the blue dotted line is OpenL3 (512 dimensions) for both image and audio without translation. We show cross-modal retrieval results of audio to image (left), and image to audio (right).}
\label{fig:cross_modal}
\end{figure*}

\section{Experimental Design}\label{sec:experi}

\subsection{Dataset}
\label{sec:dataset}

\begin{table}[h]
\centering
\begin{tabular}{c|c|c}
\hline
Subset & Stage  & \# Samples \\
\hline
Translation & Train translation & 130k \\
Cross-modal & Evaluate translation & 10k \\
\hline
\multirow{2}{*}{Classification} & Train classifier & 16.2k \\
    & Evaluate classifier & 1.8k \\
\end{tabular}
\caption{Overview of subsets used for training and evaluation.}
\label{tab:subset}
\end{table}

We use non-overlapping subsets of AudioSet \cite{audioset} for the cross-modal experiments, the training of the translation model and the classifier. AudioSet is a multi-modal dataset containing YouTube videos with weak audio labels for a diverse set of real-world situations. We follow \cite{arandjelovic2018objects} and \cite{cramer2019look} by getting samples labeled at least with a descendant of "Musical instrument", "Singing" and "Tools". We then carefully split the dataset into three subsets, one for evaluating the pre-trained embedding combinations and select the best pair, another for training the translation model, and the last one for the downstream musical instrument classification task. From all qualified videos, we sample 1 second audio and one video frame as image within the  second period. The assumption is that image and audio from roughly the same timestamp contain highly related semantic content. For evaluating the cross-modal retrieval experiments, we use a total 10k image/audio pairs. We call this subset the \textit{cross-modal-subset}. We use 130k pairs to train the translation model, which is roughly half the amount of data used to train end-to-end models in \cite{arandjelovic2018objects}. We call this the \textit{translation-subset} as shown in Table \ref{tab:subset}.

For the classification task, we carefully curated samples from 18 classes. Our categories include mapping from "Violin, fiddle" to "violin", "Choir" to "voice" and both "Drum" and "Drum kit" to "drums", and the remaining are "accordion", "banjo", "cello", "clarinet", "flute", "guitar", "mandolin", "organ", "piano", "saxophone", "synthesizer", "trombone", "trumpet", "ukulele", "cymbals". We manually audited the quality of the test set removing irrelevant samples (e.g. those labeled by piano but with image of an album cover with no piano on it) until we had 1,000 samples per instrument. Having a balanced dataset for the training of the classifier is important to prevent issues at this stage interfering with the assessment of the embeddings performance. 
We formulate the classification problem as a multi-class problem, where samples labeled only once from the above categories are selected. The result classification-subset consists of a balanced dataset with 16200 training samples and 1800 testing samples (10\% split). We call it the \textit{classification-subset}.

\subsection{Model implementation}

For ResNet and VGG image embeddings, we use the \textit{preprocess} function provided from keras to normalize the pixel values. And we follow the pre- and post-processing steps of VGGish and YamNet\footnote{\url{https://github.com/tensorflow/models/tree/master/research/audioset}}. For training the translation model, we do not use labels. Instead we randomly sample batches of size 4096 from translation-subset, extract pre-trained embeddings from both modalities, train a 2 layer MLP with both input dimensions as original pre-trained embeddings, 256 middle dimension, and 128 output dimension, implemented with PyTorch\footnote{\url{https://pytorch.org/}}. We use pairs of both modalities sampled from the same clip as positive examples, and other samples in the same batch as negative examples, with margin value as 1.0. We optimize using Adam optimizer with learning rate as 0.001, and we apply early stopping criteria on validation loss with patience as 5 epochs. For cross-modal retrieval, we take the outputs of translation model with corresponding pre-trained embeddings of the 10k image/audio pairs from cross-modal-subset, and use all embeddings from one modality as queries to fetch top 30 closest embeddings from another modality. For training the classification model, we use a random forest classifier from scikit-learn\footnote{\url{https://scikit-learn.org/}}, with maximum depth set to 32, and 100 estimators.


\subsection{Evaluation metrics}
\label{ssec:cross-modal-retrieval}

For the evaluation of the cross-modal retrieval results we follow the setup from \cite{arandjelovic2018objects}. We use \textit{normalized discounted cumulative gain} (NDCG) score considering 30 elements. This score is a measure of ranking quality between 0 and 1 (from low to high quality), which assesses the gain of an element based on a relevance score and its position in the result list. Following \cite{arandjelovic2018objects}, we use the relevance $r = C - d$, where $d$ is the distance in the taxonomy graph between two labels in the AudioSet ontology, $C = 21$ being the maximum distance. 
As the AudioSet ontology is defined, the top labels (e.g. "Music", "Musical instrument", "Tools", "Singing") are included in computing the relevance, which make most of the example relevant since most of them convey one of those labels.\footnote{See \url{https://research.google.com/audioset/ontology/index.html} for further details.} Therefore, we removed those top labels while computing the NDCG. We report the results of audio-to-image and image-to-audio retrieval.

For the evaluation of the classification results we use the macro F-measure or F1 score. Finally to assess the structural properties of the embedding spaces generated by the translation or the PCA projections we compute inter-cluster distances from the clusters of the different instrument classes and modality pairs before classification. For that, we take the test split of the classification-subset, compute average (centroids) of all the projected embeddings (PCA or translated) with the same modality and same instrument labels, and then compute pair-wise distance among modality/instrument clusters as the inter-cluster distance.

\section{Results and discussions}
\subsection{What combination of pre-trained embeddings?}

\begin{figure*}[ht]
\centering
\includegraphics[width=\linewidth]{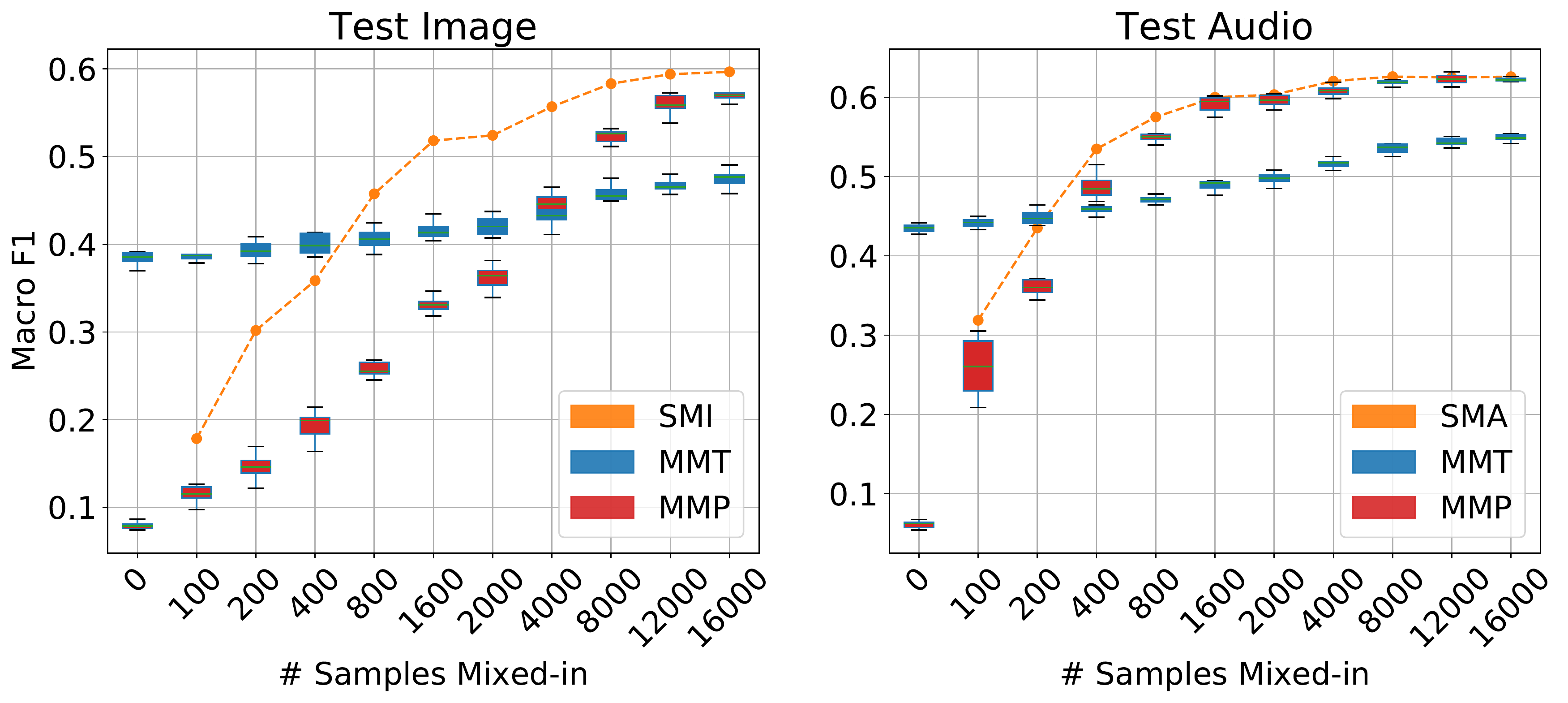}
\caption{Instrument classification results in different modalities: image (left) and audio (right). The orange dashed line is a SM classifier trained and tested with same modality, image (I) or audio (A). Blue (translated) and red (PCA) boxes are MM classifiers trained with data from different modality to the test set, with the x-axis indicating the number of samples from the target modality mixed-in in during training (e.g. when testing on image, MMT and MMP are trained with audio, and image data is mixed in by bits).}
\label{fig:train_no}
\end{figure*}

In this experiment we would like to determine which is the best audio-visual embeddings combination. We will have to simultaneously answer whether we are able to learn meaningful joint-embeddings from this data using translation, or if translation of pre-trained embeddings does not work at all.

To do so, we take all combination of audio and image embeddings and train the translation model with them, obtaining a total of nine separated translation models, i.e. nine mappings to joint embedding spaces. We then evaluate them using cross-modal retrieval in the cross-modal-subset explained in Section \ref{sec:dataset}. The NDCG scores of different configurations are depicted in Figure \ref{fig:cross_modal}, where both audio to image, and image to audio are shown. Following the ideas in \cite{arandjelovic2018objects}, we use two baselines: random (red dotted line in Figure \ref{fig:cross_modal}) which means randomly ordering the embeddings and get the first 30, and the OpenL3 (blue dotted line) image and audio embeddings both with 512 dimension \footnote{Note that the OpenL3 implementation allows for multiple output dimensions, and we choose 512 here for both embeddings to be comparable.}, used for  retrieval directly without translation. 

The first observation is that the relative difference in NDCG scores between the baselines and our best performing model are comparable to those shown in \cite{arandjelovic2018objects},
which is promising because it means that the translation model is effectively learning to relate the embeddings. Also unlike the systems in \cite{arandjelovic2018objects} which were trained from scratch, we obtained our joint embedding by translating pre-trained embeddings, and obtained competitive results. The random baseline  performs better than OpenL3 for audio to image retrieval, which is consistent with the results reported in \cite{arandjelovic2018objects}.

We observe a big gap in performance in all combinations that include the OpenL3 image embedding, which can be partially explained by the fact that VGG and ResNet greatly outperform that embedding in image classification downstream tasks, and thus more expressive embeddings would be better candidates for translation.

Overall, the combination of YamNet and ResNet performs the best across all configurations. We checked that is the case for the classification performance as well, therefore, we discuss only YamNet and ResNet results in the rest of the experiments.

\subsection{How does translation affect performance?}




We want to understand how translation affects the performance of a classifier in comparison to its multi-modal non-translated and single-modality counterparts. For that we compare their performance in the classification task, by training the classifier using embeddings from one modality and testing with embeddings from the other, and by adding batches of the training modality with balanced number of instrument classes by bits. We use the classification-subset for this experiment. The results of this process are shown in Figure \ref{fig:train_no}, where the macro F1 scores are reported for a test set of only images (left) and only audio (right). 

\noindent \textbf{No data from target modality}. First, we discuss the results in the $0$ point of the x-axis, corresponding to the performance of the classifiers without any data from the target modality (e.g. when testing in image, only training the MM classifiers with audio). We see a similar and expected behaviour in both modalities: the MMP classifier is guessing some classes right (very little), probably exploiting some unintended relations between YamNet and ResNet embeddings after PCA, and the MMT classifier clearly outperforms the others, being able to achieve almost 70\% of the best performance already. This confirms what we saw from the cross-modal retrieval examples, that the translation is doing a meaningful mapping, and further this allows to learn from one modality and test in another in a zero-shot fashion. 

\noindent \textbf{Adding data from target modality.} However, for an ideal translation, image and audio embeddings would be interchangeable. That means that the performance of MMT without seeing any embedding from the target modality or after seeing all of them should be the same (since no \textit{new} data would be added to the classifier). And so, the blue curve we see in Figure \ref{fig:train_no} with a small slope should flat at the maximum performance independently of the test data we add in. The fact that the performance of MMT increases by adding this data is showing that the translation failed in combining \textit{some} meaningful information. And this makes the MMT classifier's performance to fall behind when all the data from both modalities are available (point 16000 in the x-axis of Figure \ref{fig:train_no}). 

Observing the performance of MMP, we also see that with the right amount of data and without translation, the classifier is able to learn the mapping between embeddings and classify the instruments correctly. This is an interesting result since it implies that for specific tasks with available labeled data from both (or multiple) modalities, it is enough to train a classifier to learn to deal with different modalities all together and be able to work with whatever modality is available at inference time with almost the same performance than a classifier fully dedicated to one modality.

\begin{figure*}[ht]
\centering
\includegraphics[width=\linewidth]{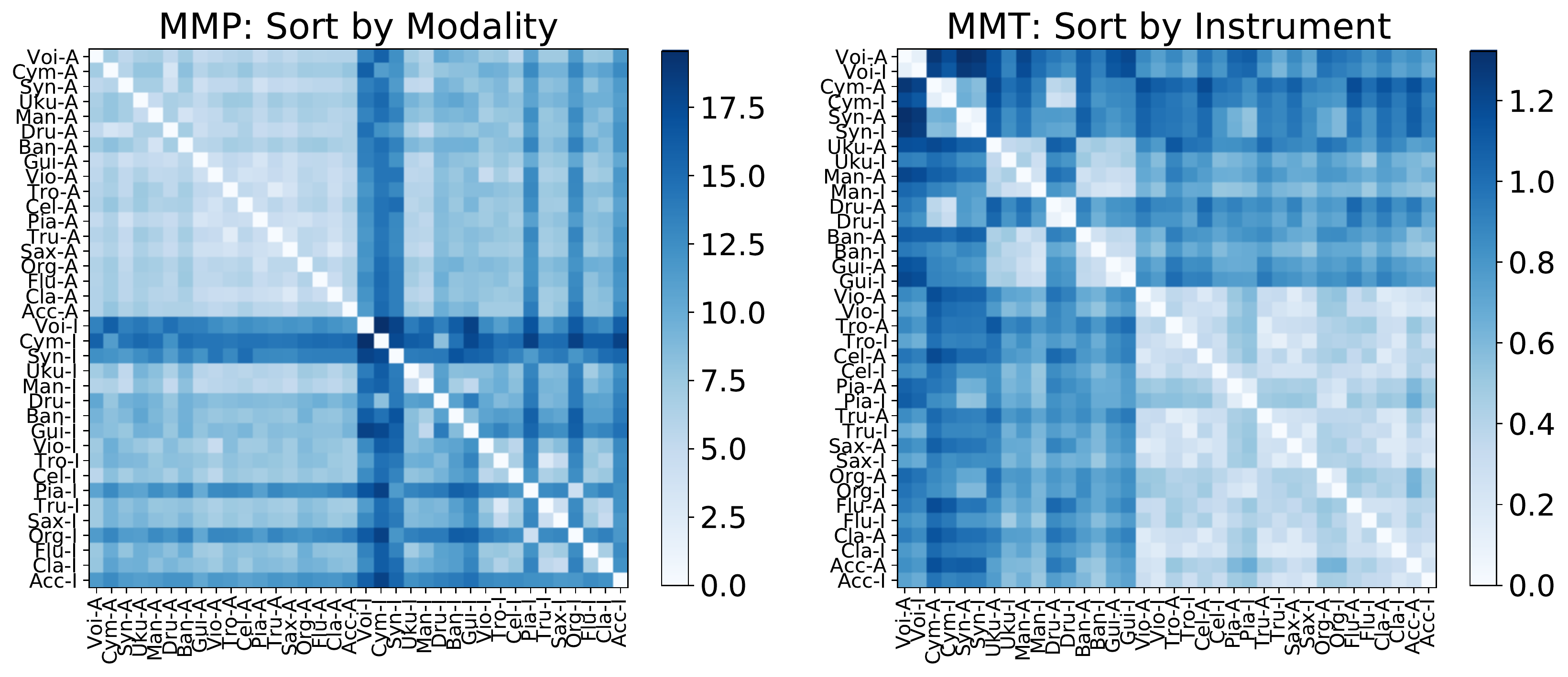}
\caption{Pair-wise distance of inter-cluster centroid for modality and instrument classes. On the left we have PCA sorted first by modality. On the right we have translated sorted first by instrument class. Note: This is results of classification-subset but before training the classifier.}
\label{fig:pca_translate_sorted}
\end{figure*}

\subsection{What is translation doing?}

We want to understand what is happening during translation that the MMT classifier is struggling to keep up with the others when enough data from both modalities is available, and why it does not reach best performance starting from the zero-shot setting. We compare the structure of the non-translated and translated embeddings before feeding them into the classifier. In particular we measure the distance between the cluster centroids of the different clusters of classes in each setting. Figure \ref{fig:pca_translate_sorted} shows the pair-wise distance for the different modalities and instrument classes. On the left we see the non-translated embeddings sorted by modality, and on the right we see translated embeddings sorted by instrument class. The two figures show that the embedding spaces are indeed different, and that the translation is structuring and bringing together the audio and image embeddings of the same class (shown as small 2x2 squares on the diagonal), i.e. grouping the embeddings by concepts. This makes sense and explains why the translation works in the zero-shot setting, and is interesting considering that the translation layer is trained in an unsupervised manner.

However, there are noticeable small distance square blocks in both images: the one in the left on the top left, shows that the audio embeddings are closer to each other, which is an artifact of YamNet embeddings. This is not what happens with the ResNet embeddings, which after the PCA projection are sometimes closer to other image embeddings but also sometimes closer to audio embeddings. An exception to this are the image embeddings for voice, cymbals and synthesizer, which are very different from all the other embeddings.

\begin{figure}[!ht]
\centering
\includegraphics[width=\linewidth]{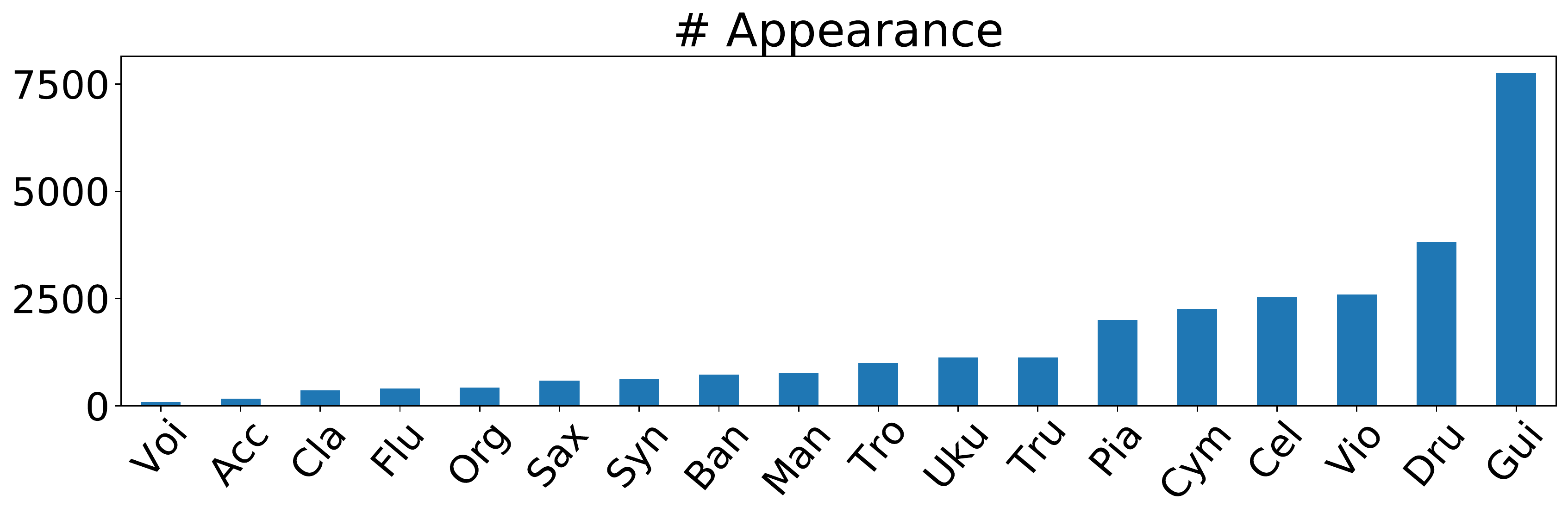}
\caption{Number of samples containing classification instrument labels in translation-subset. We show here only the classes under study in the classification task.}
\label{fig:cooccurrence}
\end{figure}

The other big block where embeddings are grouped together, in the right image of Figure \ref{fig:pca_translate_sorted}, shows that the translation is bringing together some classes (e.g. accordian, clarinet, flute, organ, saxophone, etc.) that should not be blend together, and the overall distances in the translated embedding space are smaller than in the non-translated one. Observing the class distribution of the data used for training the translation model in Figure \ref{fig:cooccurrence} (note the labels were not use in the training, only here for the analysis), we observe it is skewed and the classes with fewer number of instances correlate with most of the confused ones in the translated embeddings. We think that this is probably causing the classification performance of MMT to drop with respect to MMP and SM. The exception is voice, which we speculate has to do with an effect from the ResNet embedding which is very different from all other embeddings and we suspect that helped the translated cluster to be sufficiently different as well.


\begin{figure}[!ht]
\centering
\includegraphics[width=\linewidth]{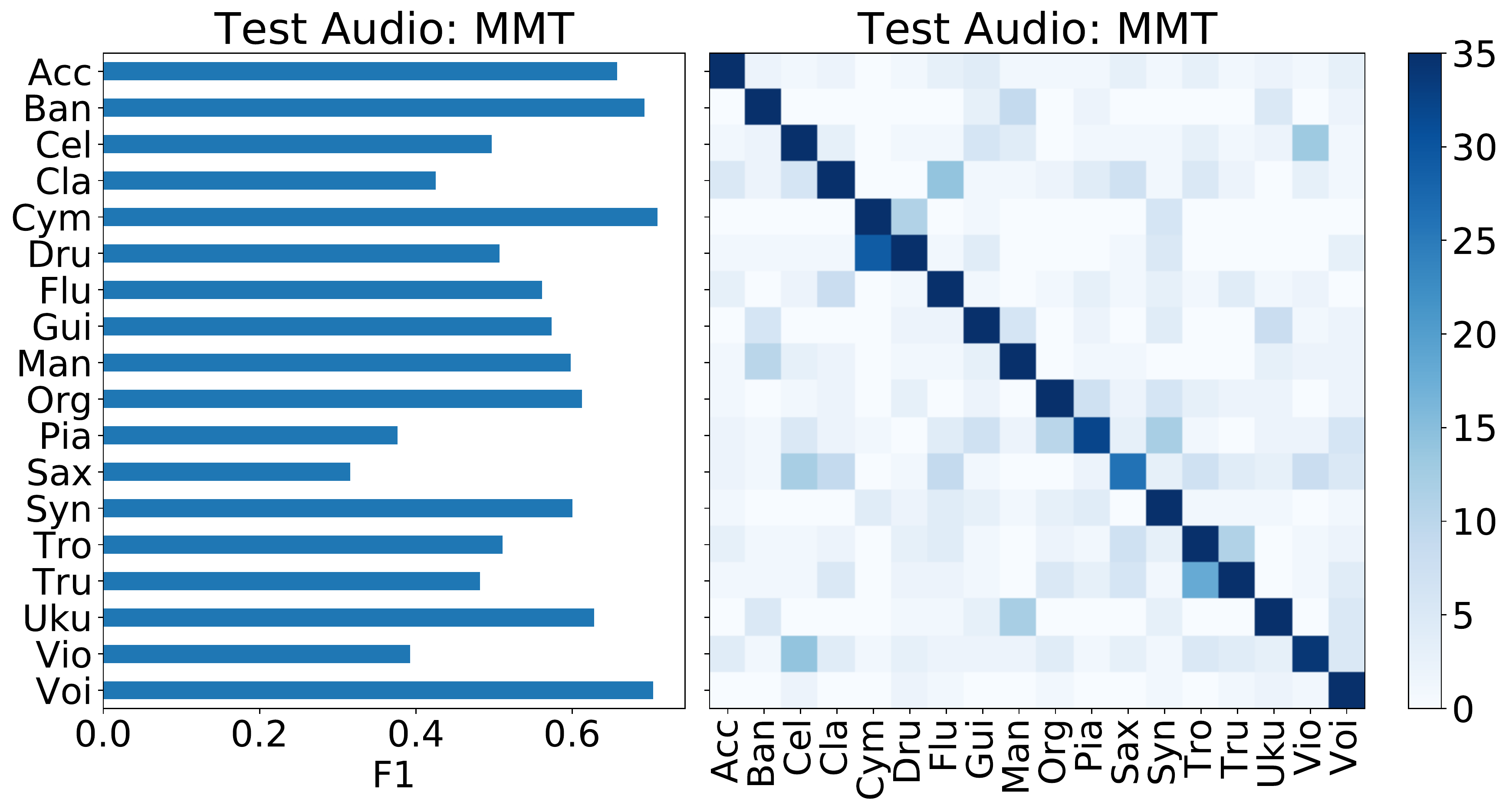}
\includegraphics[width=\linewidth]{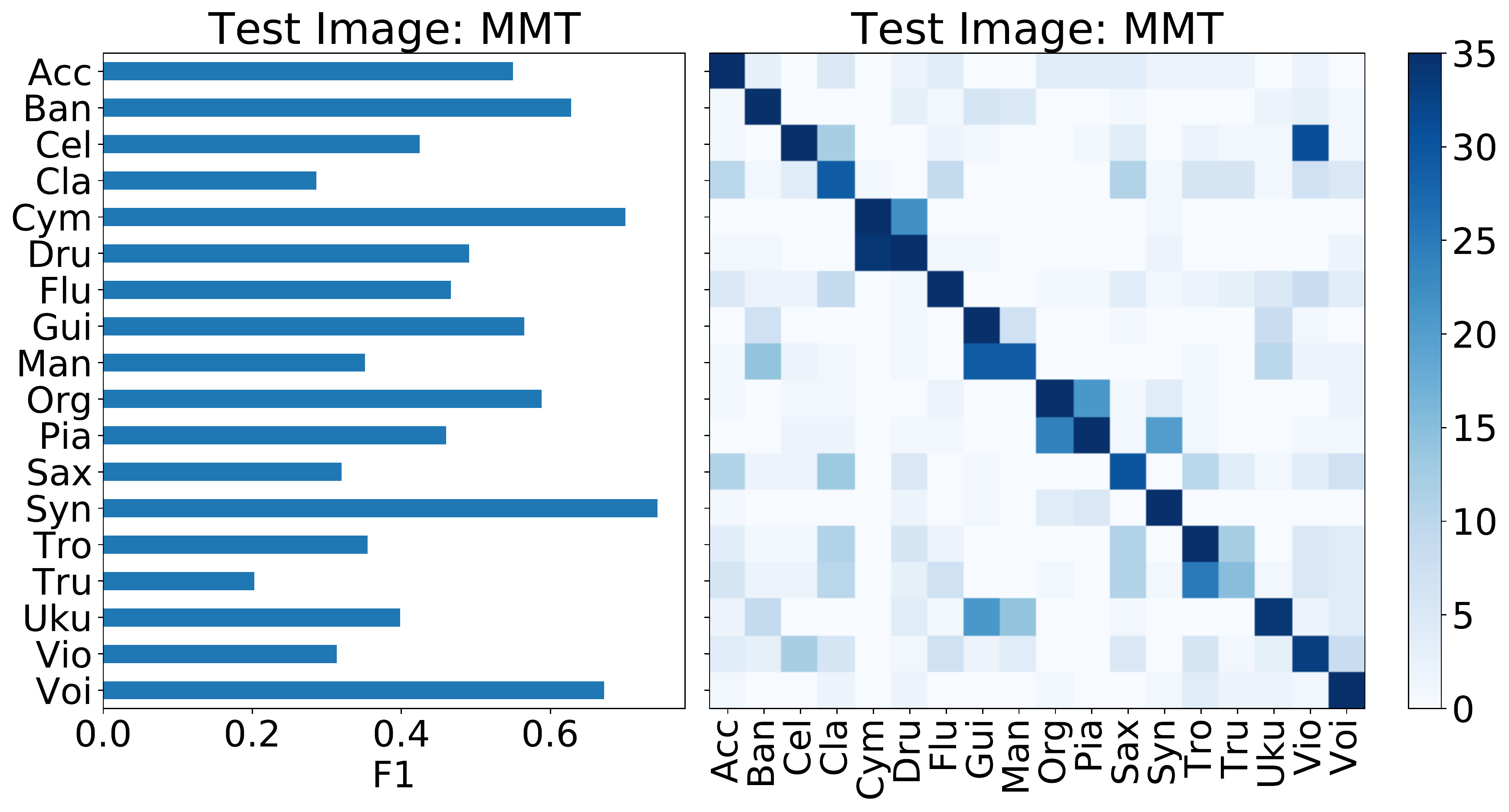}
\caption{Classification results: training with all data using both modalities and testing in audio (top) and image (bottom). Per instrument F1 on the left, confusion matrix on the right.}
\label{fig:cm_test_image}
\end{figure}

To see the correlation between our observations in the embedding space and the classification performance, we look at the per instrument F1 and confusion matrices of the MMT classifier, using all data from both modalities as shown in Figure \ref{fig:cm_test_image}. In each figure we show the per instrument F1 on the left, and the confusion matrix of MMT on the right. Looking at the per instrument F1 on the left, we can also observe the correlations between less performant instrument classes with smaller distances in Figure \ref{fig:pca_translate_sorted} and number of fewer samples in Figure \ref{fig:cooccurrence}. Looking at the confusion matrix of MMT on the right, we see that there are common mistakes of cymbal vs drum (both modalities), trombone vs trumpet (both modalities), guitar vs mandolin (mostly image), and organ vs piano (mostly image), which make sense because of the acoustic or visual similarity of those instruments in each modality. We observe a similar trend in the confusions made by the MMP classifier, but in a lesser extent (which explains the better performance).

To sum up, we believe that the bias of label distribution in the data we used for training translation is the main cause of the performance drop in classification, and this is a trade-off of self-supervised learning without using the labels. We plan to explore in the future unsupervised methods for sample selection to balance the training set used for the translation model, such as determinant point processes \cite{kulesza2012determinantal}.

\section{Conclusions and future work}\label{sec:conclusions}


In this work we propose and investigate modality-agnostic representations for music classification. We first present a study on different combinations of pre-trained audio and image embeddings to determine the best configuration to obtain modality-agnostic representations via cross-modal evaluation. We then use this representation to train instrument classifiers, comparing with non-translated and single-modality baselines. We show promising results as well as interesting potential applications using data from one modality to train and another modality to test with reasonable performance (almost 70\% of best performing system in a zero-shot setting). We also investigate how biases in the training data used for the translation affect the classification performance.

For future work, we are interested in exploring sampling methods \cite{wu2017sampling, won2021multimodal} that could help balance the training set to obtain a more unbiased translation model, which from our analysis could lead to better performance. Also, we are interested in exploring the joint training of the translation and classification models, instead of the sequential method proposed in this paper. Furthermore, we think that exploring novel loss functions specifically for multi-modal data will also be an interesting direction as most of the current contrastive methods are applied to single modality. This work presented first steps and analysis towards the use of modality-agnostic representations in music, which we consider to be a promising idea in the context of MIR since it allows the use of data from different datasets and modalities in a flexible way, relaxing concerns about data scarcity and other data-availability related issues.


\begin{acknowledgments}
This work is partially supported by the National Science Foundation award \#1544753. Magdalena Fuentes is a faculty fellow in the NYU
Provost’s Postdoctoral Fellowship Program at the NYU
Center for Urban Science and Progress and Music and Audio Research Laboratory.
\end{acknowledgments}
	
\bibliography{smc2021bib}
	
\end{document}